\documentclass[12pt,prd,showpacs,tightenlines,nofootinbib]{revtex4}
\usepackage{bm}
\usepackage{graphics}
\usepackage{rotating}
\usepackage{epsfig}
\begin{document}
\title{\begin{flushright}{\rm\normalsize HU-EP-06/04}\end{flushright}
Relativistic treatment of the decay constants of light
and heavy mesons}
\author{D. Ebert}
\affiliation{Institut f\"ur Physik, Humboldt--Universit\"at zu Berlin,
Newtonstr. 15, D-12489  Berlin, Germany}
\author{R. N. Faustov}
\author{V. O. Galkin}
\affiliation{Institut f\"ur Physik, Humboldt--Universit\"at zu Berlin,
Newtonstr. 15, D-12489 Berlin, Germany}
\affiliation{Dorodnicyn Computing Centre, Russian Academy of Sciences,
  Vavilov Str. 40, 119991 Moscow, Russia}

\begin{abstract}
Novel relativistic expressions are used to calculate the weak
decay constants of pseudoscalar and vector mesons within the
constituent quark model. Meson wave functions satisfy the
quasipotential equation with the complete relativistic potential.
New contributions, coming from the negative-energy quark states,
are substantial for the light mesons, significantly decrease the
values of their decay constants and, thus, bring  them into
agreement with experiment. For heavy-light mesons these
contribution are much less pronounced, but permit to reduce
uncertainties of the predicted decay constants. Their values agree
with the results of lattice calculations and experimental data.

\end{abstract}


\maketitle

The weak decay constants of pseudoscalar and vector mesons belong to 
their most important characteristics, which enter in various decay rates. 
Many efforts were undertaken to calculate these constants within lattice
QCD (both quenched and unquenched) \cite{akr,ak,milc,ml,hpqcd,chllc}, QCD
sum rules \cite{nar,ps,jl},  
and constituent quark models \cite{gi,mr,k,gmf,efgdc,ckwn,hjd}. At present,
the decay constants of light mesons are measured with high precision, 
while in the heavy-light meson sector only $D$ and $D_s$ meson decay
constants are available with rather large errors \cite{pdg}. Recently,
a relatively precise experimental value for the 
$D$ meson decay constant was presented by the CLEO Collaboration \cite{cleo}. 
Therefore it is actual to reconsider the meson decay 
constants treating  quarks, composing the meson, in a consistently
relativistic way. 
Such procedure was formulated and successfully applied 
for light mesons in the papers \cite{lmm}. In this letter we evaluate new
contributions to relativistic expressions for the meson decay
constants coming from the negative-energy quark states both for light
and heavy-light mesons. We use the meson wave functions satisfying the
quasipotential equation with the complete 
relativistic potential in order to obtain new, 
more accurate predictions for the meson decay constants.

  In the quasipotential approach a meson is described by the wave
function of the bound quark-antiquark state~\cite{efg}, which satisfies the
quasipotential equation  of the Schr\"odinger type
\begin{equation}
\label{quas}
{\left(\frac{b^2(M)}{2\mu_{R}}-\frac{{\bf
p}^2}{2\mu_{R}}\right)\Psi_{M}({\bf p})} =\int\frac{d^3 q}{(2\pi)^3}
 V({\bf p,q};M)\Psi_{M}({\bf q}),
\end{equation}
where the relativistic reduced mass is
\begin{equation}
\mu_{R}=\frac{E_1E_2}{E_1+E_2}=\frac{M^4-(m^2_1-m^2_2)^2}{4M^3},
\end{equation}
and $E_1$, $E_2$ are given by
\begin{equation}
\label{ee}
E_1=\frac{M^2-m_2^2+m_1^2}{2M}, \quad E_2=\frac{M^2-m_1^2+m_2^2}{2M}.
\end{equation}
Here $M=E_1+E_2$ is the meson mass, $m_{1,2}$ are the quark masses,
and ${\bf p}$ is their relative momentum.  
In the center-of-mass system the relative momentum squared on mass shell 
reads
\begin{equation}
{b^2(M) }
=\frac{[M^2-(m_1+m_2)^2][M^2-(m_1-m_2)^2]}{4M^2}.
\end{equation}

The kernel 
$V({\bf p,q};M)$ in Eq.~(\ref{quas}) is the quasipotential operator of
the quark-antiquark interaction. It is constructed with the help of the
off-mass-shell scattering amplitude, projected onto the positive
energy states. 
Constructing the quasipotential of the quark-antiquark interaction, 
we have assumed that the effective
interaction is the sum of the usual one-gluon exchange term with the mixture
of long-range vector and scalar linear confining potentials, where
the vector confining potential
contains the Pauli interaction. The quasipotential is then defined by
\footnote{In our notation, where strong annihilation processes are neglected,
 antiparticles are described by usual spinors  taking into account the
 proper quark charges.} 
  \begin{equation}
\label{qpot}
V({\bf p,q};M)=\bar{u}_1({\bf p})\bar{u}_2(-{\bf p}){\mathcal V}({\bf p}, 
{\bf q};M)u_1({\bf q})u_2(-{\bf q}),
\end{equation}
with
$${\mathcal V}({\bf p},{\bf q};M)
\equiv{\mathcal V}({\bf p}-{\bf q})=\frac{4}{3}\alpha_sD_{ \mu\nu}({\bf
k})\gamma_1^{\mu}\gamma_2^{\nu}
+V^V_{\rm conf}({\bf k})\Gamma_1^{\mu}
\Gamma_{2;\mu}+V^S_{\rm conf}({\bf k}),$$
where $\alpha_s$ is the QCD coupling constant, $D_{\mu\nu}$ is the
gluon propagator in the Coulomb gauge and
${\bf k=p-q}$; $\gamma_{\mu}$ and $u({\bf p})$ are Dirac matrices and spinors.
The effective long-range vector vertex is given by
\begin{equation}
\label{kappa}
\Gamma_{\mu}({\bf k})=\gamma_{\mu}+
\frac{i\kappa}{2m}\sigma_{\mu\nu}k^{\nu},
\end{equation}
where $\kappa$ is the Pauli interaction constant characterizing the
anomalous chromomagnetic moment of quarks. Vector and
scalar confining potentials in the nonrelativistic limit reduce to
\begin{eqnarray}
\label{vlin}
V^V_{\rm conf}(r)&=&(1-\varepsilon)(Ar+B),\nonumber\\ 
V^S_{\rm conf}(r)& =&\varepsilon (Ar+B),
\end{eqnarray}
reproducing 
\begin{equation}
\label{nr}
V_{\rm conf}(r)=V^S_{\rm conf}(r)+V^V_{\rm conf}(r)=Ar+B,
\end{equation}
where $\varepsilon$ is the mixing coefficient. 

All the model parameters have the same values as in our previous
papers \cite{egf,efg}.
The constituent quark masses $m_u=m_d=0.33$ GeV, $m_s=0.5$ GeV,
$m_c=1.55$ GeV, $m_b=4.88$ GeV and
the parameters of the linear potential $A=0.18$ GeV$^2$ and $B=-0.3$ GeV
have the usual values of quark models.  The values of the mixing
coefficient of vector and scalar confining potentials $\varepsilon=-1$
and the universal Pauli interaction constant $\kappa=-1$ are specific
for our model. 

The quasipotential (\ref{qpot}) can  be used for arbitrary quark
masses.  The substitution 
of the Dirac spinors  into (\ref{qpot}) results in an extremely
nonlocal potential in the configuration space. Clearly, it is very hard to 
deal with such potentials without any additional transformations.
 In oder to simplify the relativistic $q\bar q$ potential, we make the
following replacement in the Dirac spinors \cite{egf,lmm}:
\begin{equation}
  \label{eq:sub}
  \epsilon_{1,2}(p)=\sqrt{m_{1,2}^2+{\bf p}^2} \to E_{1,2}.
\end{equation}
This substitution makes the Fourier transformation of the potential
(\ref{qpot}) local, but the resulting relativistic potential becomes
dependent on the meson mass in a very complicated nonlinear way. 
We consider only the meson ground
states, which further simplifies our analysis, since all terms
containing orbital momentum vanish. The detailed expressions for the
relativistic quark potential can be found in Ref.~\cite{lmm}. Here we
use these formulas for the calculation of the ground state meson masses. 
We solve numerically the quasipotential equation with the local
fully relativistic potential, which includes both spin-independent and
spin-dependent parts. As a result we get the relativistic wave
functions of the ground state mesons which depend
nonperturbatively on the meson spin  (i.e. the pseudoscalar and vector
meson wave functions are different). These wave functions are used
below for calculating the decay constants of light and heavy mesons.
The obtained masses of the pseudoscalar and vector 
mesons are given in Table~\ref{tab:mass} in comparison with the
experimental data \cite{pdg}. The overall good agreement of our
predictions with experiment is found.

\begin{table}
  \caption{Masses of the ground state light and heavy-light mesons (in MeV).}
  \label{tab:mass}
\begin{ruledtabular}
\begin{tabular}{ccc}
Meson& $M^{\rm theor}$ & $M^{\rm exp}$ PDG \cite{pdg}\\ 
\hline
$\pi$& 154 &139.57\\
$\rho$ & 776 & 775.8(5)\\
$K$ & 482& 493.677(16)\\
$K^*$& 897& 891.66(26)\\
$\phi$& 1038& 1019.46(2)\\
$D$& 1872 & 1869.4(5)\\
$D^*$& 2009 & 2010.0(5)\\
$D_s$& 1967 & 1968.3(5)\\
$D_s^*$& 2112 & 2112.1(7)\\
$B$ & 5275 & 5279.0(5)\\
$B^*$ & 5326 & 5325.0(6)\\
$B_s$ & 5362 & 5369.6(2.4)\\
$B_s^*$ & 5414 & 5416.6(3.5)
  \end{tabular}
\end{ruledtabular}
\end{table}

\begin{figure}
  \centering
  \includegraphics[width=16cm]{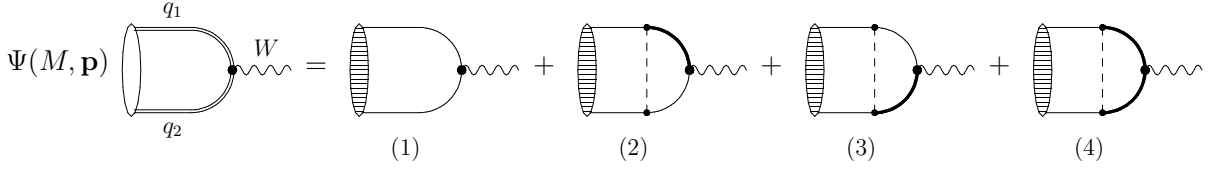}
  \caption{Weak annihilation diagram of the  meson. Solid and bold
lines denote the  positive- and negative-energy part of the quark propagator,
respectively. Dashed lines represent the interaction operator
${\mathcal V}$. Dashed ovals depict the projected wave function
$\Psi_{M\, {\bf K}}({\bf p})$.} 
  \label{fig:diag}
\end{figure}

The decay constants $f_P$ and $f_V$ of the pseudoscalar ($P$) and
vector ($V$) mesons parameterize the matrix elements of the weak
current $J^W_\mu=\bar q_1{\cal J}^W_\mu q_2=\bar q_1\gamma_\mu(1-\gamma_5)q_2$
between the corresponding meson and vacuum states. They are defined by  
\begin{eqnarray}
  \label{eq:dc}
  \left<0|\bar q_1 \gamma^\mu\gamma_5 q_2|P({\bf K})\right>&=& i f_P
  K^\mu,\\ 
\left<0|\bar q_1 \gamma^\mu q_2|V({\bf K},\varepsilon)\right>&=& f_V
  M_V \varepsilon^\mu,
\end{eqnarray}
where ${\bf K}$ is the meson momentum, $\varepsilon^\mu$ and $M_V$ are
the polarization vector and mass of the vector meson. This matrix
element can be expressed through the two-particle Bethe-Salpeter wave
function $\Psi(M,p)$ in the quark loop integral (see Fig.~\ref{fig:diag})
\begin{equation}
  \label{eq:bs}
  \left<0| J^W_\mu |M({\bf K})\right>=\int\frac{d^4
    p}{(2\pi)^4}
{\rm Tr}\left\{\gamma_\mu(1-\gamma_5)\Psi(M,p)\right\},
\end{equation}
where the trace is taken over spin indices. Integration 
over $p^0$ in Eq.~(\ref{eq:bs}) allows one to pass to the Fourier 
transform of the single-time
wave function in the meson rest frame
\begin{equation}
  \label{eq:stw}
  \Psi(M,{\bf p})=\int\frac{dp^0}{2\pi}\Psi(M,p).
\end{equation}
This wave function contains both positive- and negative-energy
quark states. Since in the quasipotential approach we use the
wave function $\Psi_{M\, {\bf K}}({\bf p})$  projected onto the
positive-energy states, it is necessary to include additional terms
which account for the contributions of negative-energy intermediate
states. Within perturbation theory the weak matrix element
(\ref{eq:bs}) is schematically presented in 
Fig.~\ref{fig:diag}. The first diagram in the right hand side
corresponds to the simple replacing of the wave function
(\ref{eq:stw})  $\Psi(M,{\bf p})$ by the projected one
$\Psi_{M\, {\bf K}}({\bf p})$.\footnote{The contributions with the exchange by the
  effective interaction potential ${\mathcal V}$ which contain only
  positive-energy intermediate states are automatically accounted for
  by the wave function itself.}  The second and third diagrams account for
negative-energy contributions to the first and second quark propagators,
respectively. The last diagram corresponds to negative-energy
contributions from both quark propagators. 

 Thus in the quasipotential approach
this decay amplitude has the form
\begin{eqnarray}
  \label{eq:qpd}
\left<0|J^W_\mu |M({\bf K})\right>&=&\sqrt{2M}\Biggl\{\int\frac{d^3
  p}{(2\pi)^3} \bar u_1({\bf p}_1){\cal J}^W_\mu u_2({\bf p}_2)
 \Psi_{M\, {\bf K}}({\bf
  p})+\Biggl[\int\frac{d^3 p d^3 p'}{(2\pi)^6}\bar u_1({\bf p}_1)\Gamma_1\cr 
&&\!\!\!\!\times
\frac{\Lambda^{(-)}_1({\bf p}_1')\gamma^0{\cal J}^W_\mu
\Lambda^{(+)}_2({\bf p}_2')\gamma^0}
{M+\epsilon_1(p')-\epsilon_2(p')}\Gamma_2 u_2({\bf p}_2)
\tilde V({\bf p}-{\bf p}')
\Psi_{M\, {\bf K}}({\bf p})
+(1\leftrightarrow 2)\Biggr]\cr
&&\!\!\!\!\!\!\!\!\!+\int\frac{d^3 p d^3 p'}{(2\pi)^6}\bar u_1({\bf p}_1)\Gamma_1
\frac{\Lambda^{(-)}_1({\bf p}_1')\gamma^0{\cal J}^W_\mu
\Lambda^{(-)}_2({\bf p}_2')\gamma^0}
{M+\epsilon_1(p')+\epsilon_2(p')}\Gamma_2 u_2({\bf p}_2)
\tilde V({\bf p}-{\bf p}')
\Psi_{M\, {\bf K}}({\bf
  p})\Biggr\},\cr&&\!\!\!
\end{eqnarray}
where ${\bf p}_{1,2}^{(')}={\bf K}/2\pm {\bf p}^{(')}$;
$\epsilon(p)=\sqrt{{\bf p}^2+m^2}$; matrices
$\Gamma_{1,2}$ denote the Dirac structure of the interaction 
potential (\ref{qpot}) for the first and second quark, respectively, and thus
$\Gamma_1\Gamma_2\tilde V({\bf p}-{\bf p}')={\mathcal V}({\bf p}-{\bf p}')$. 
The factor $\sqrt{2M}$ follows
from the normalization of the quasipotential wave function. The
positive- and negative-energy projectors have standard definition
\[
\Lambda^{(\pm)}({\bf p})={\epsilon(p)\pm\bigl( m\gamma ^0+\gamma^0(
\bm{\gamma}{\bf p})\bigr) \over 2\epsilon (p)}.
\]  
The ground-state wave function  in the rest frame
of the decaying meson $\Psi_M({\bf p})\equiv\Psi_{M\, {\bf 0}}({\bf p})$
can be expressed through a product of radial $\Phi_M({\bf p})$,
spin $\chi_{ss'}$ and colour $\phi_{q_1q_2}$ wave functions
\begin{equation}
  \label{eq:wff}
  \Psi_M({\bf p})=\Phi_M({\bf p})\chi_{ss'}\phi_{q_1q_2}.
\end{equation}

Now the decay constants can be presented in the following form
\begin{equation}
  \label{eq:fpe}
  f_{P,V}=f_{P,V}^{(1)}+f_{P,V}^{(2+3)}+f_{P,V}^{(4)},
\end{equation}
where the terms on the right hand side originate from the corresponding
diagrams in Fig.~\ref{fig:diag} and parameterize respective terms in
Eq.~(\ref{eq:qpd}). In the literature \cite{gi,efgdc} usually only
the first term is taken 
into account, since it provides the nonrelativistic limit, while
other terms give only relativistic corrections and thus vanish in
this limit. Such approximation can be justified for  mesons
containing only heavy quarks. However,  as it will be shown below, for
mesons with light quarks, especially for light mesons, other terms
become equally important and their account is 
crucial for getting the results in agreement with experimental data. 

The matrix element (\ref{eq:qpd}) and thus the decay constants can be
calculated in an arbitrary frame and from any component of the weak
current \cite{gmf}. Such calculation can 
be most easily performed in the rest frame of the decaying meson from the
zero component of the current. The same results will be obtained from the
vector component; however, this calculation is more cumbersome, since
then the rest frame cannot be used and, thus, it is necessary to take
into account the relativistic transformation of the meson wave function
from the rest frame to the moving one with the momentum ${\bf K}$. It
is also possible to perform calculations in an explicitly covariant
way using methods proposed in \cite{efgms}.   

The resulting expressions for decay constants are given by
\begin{eqnarray}
  \label{eq:fpv1}
  f^{(1)}_{P,V}&=&\sqrt{\frac{12}{M}}\int \frac{d^3
  p}{(2\pi)^3}\left(\frac{\epsilon_1(p)+m_1}{2\epsilon_1(p)}\right)^{1/2}
  \left(\frac{\epsilon_2(p)+m_2}{2\epsilon_2(p)}\right)^{1/2}
 \cr
&&\times \left\{ 1
  +\lambda_{P,V}\,\frac{{\bf p}^2}{[\epsilon_1(p)+m_1][\epsilon_2(p)+m_2]}\right\}
  \Phi_{P,V}({\bf p}),
\end{eqnarray}
\begin{eqnarray}
  \label{eq:fpv3}
  f^{(2+3)}_{P,V}&=&\sqrt{\frac{12}{M}}\int \frac{d^3
  p}{(2\pi)^3}\left(\frac{\epsilon_1(p)+m_1}{2\epsilon_1(p)}\right)^{1/2}
  \left(\frac{\epsilon_2(p)+m_2}{2\epsilon_2(p)}\right)^{1/2}\Biggl[
\frac{M-\epsilon_1(p)-\epsilon_2(p)}{M+\epsilon_1(p)-\epsilon_2(p)}\cr
&&\times
\frac{{\bf p}^2}{\epsilon_1(p)[\epsilon_1(p)+m_1]}
  \left\{1+\lambda_{P,V}\frac{\epsilon_1(p)+m_1}{\epsilon_2(p)+m_2}\right\}
+(1\leftrightarrow 2)\Biggr]
  \Phi_{P,V}({\bf p}),
\end{eqnarray}
\begin{eqnarray}
  \label{eq:fpv4}
  f^{(4)}_{P,V}&=&\sqrt{\frac{12}{M}}\int \frac{d^3
  p}{(2\pi)^3}\left(\frac{\epsilon_1(p)+m_1}{2\epsilon_1(p)}\right)^{1/2}
  \left(\frac{\epsilon_2(p)+m_2}{2\epsilon_2(p)}\right)^{1/2}
\frac{M-\epsilon_1(p)-\epsilon_2(p)}{M+\epsilon_1(p)+\epsilon_2(p)}\cr
&&\times
  \left\{-\lambda_{P,V}-\frac{{\bf p}^2}{[\epsilon_1(p)+m_1]
[\epsilon_2(p)+m_2]}\right\} \cr
&&\times
\left[\frac{(1-\varepsilon)m_1^2m_2^2}{\epsilon_1^2(p)\epsilon_2^2(p)}+
\frac{{\bf p}^2}{[\epsilon_1(p)+m_1][\epsilon_2(p)+m_2]}\right]
  \Phi_{P,V}({\bf p}),
\end{eqnarray}
with $\lambda_P=-1$ and $\lambda_V=1/3$. Here $\varepsilon$ is the
mixing coefficient of scalar and vector confining potentials
(\ref{vlin}) and the long-range anomalous chromomagnetic quark moment
$\kappa$ (\ref{kappa}) is put equal to $-1$. Note that  $f_P^{(2+3)}$
vanishes for pseudoscalar mesons with equal quark masses, such as the pion.
The positive-energy contribution (\ref{eq:fpv1}) reproduces the previously known
expressions for the decay constants \cite{gi}. The negative-energy
contributions (\ref{eq:fpv3}) and (\ref{eq:fpv4}) are new and play a
significant role for light mesons.

In the nonrelativistic limit ${\bf p}^2/m^2\to 0$ the expression 
(\ref{eq:fpv1}) for decay constants gives the well-known formula
\begin{equation}
\label{eq:fnr}
f_{P,V}^{\rm NR}=
\sqrt{\frac{12}{M_{P,V}}}\left|\Psi_{P,V}(0)\right|,
\end{equation}
where $\Psi_{P,V}(0)$ is the meson wave function at the origin
$r=0$. All other contributions vanish in the nonrelativistic limit.

\begin{table}
  \caption{Different contributions to the pseudoscalar and vector
    decay constants of light and heavy mesons (in  MeV). The notations are taken
  according to Eqs.~(\ref{eq:fpe}) and (\ref{eq:fnr}).} 
  \label{tab:dc}
\begin{ruledtabular}
\begin{tabular}{cccccc}
Constant& $f_M^{\rm NR}$&$f_M^{(1)}$& $f_M^{(2+3)}+f_M^{(4)}$
& $(f_M^{(2+3)}+f_M^{(4)})/f^{(1)}_M $&$f_M$ \\
\hline
$f_\pi$ & 1290 &515 & $-391$ &$-76\%$ &124  \\
$f_\rho$ & 490 & 402 & $-183$&$-46\%$ & 219 \\
$f_K$ & 783 & 353 & $-198$& $-56\%$ & 155 \\
$f_{K^*}$ & 508 & 410 & $-174$&$-42\%$ & 236\\
$f_\phi$ & 511 & 415 &$-170$&$-41\%$  &245 \\
$f_D$ & 376 & 275 &$-41$& $-15\%$ & 234\\
$f_{D^*}$ & 391 & 334 & $-24$ & $-7\%$ & 310\\
$f_{D_s}$ & 436 & 306 & $-38$ & $-12\%$ & 268\\
$f_{D_s^*}$ & 447 & 367 & $-52$ & $-14\%$ & 315\\
$f_B$ & 259 & 210 & $-21$ & $-10\%$ & 189\\
$f_{B^*}$ & 280 & 235 & $-16$ & $-7\%$ & 219\\
$f_{B_s}$ & 300 & 238 & $-20$ & $-8\%$ & 218\\
$f_{B_s^*}$ & 316 & 264 & $-13$ & $-5\%$ & 251
\end{tabular}
\end{ruledtabular}

\end{table}

\begin{table}
  \caption{Pseudoscalar and vector decay constants of light mesons (in
    MeV).}
  \label{tab:dce}
\begin{ruledtabular}
\begin{tabular}{ccccccccc}
Constant&this work&\cite{gi} & \cite{mr} &\cite{k} &\cite{hjd}
&Lattice \cite{ak} &Lattice \cite{milc}& Experiment \cite{pdg}\\
\hline
$f_\pi$ & 124& 180 & 131&219&138&$126.6(6.4)$ &$129.5(3.6)$  &$130.70(10)(36)$\\
$f_K$ & 155& 232& 155& 238&160&$152.0(6.1)$&$156.6(3.7)$ & $159.80(1.4)(44)$\\
$f_\rho$ & 219&220&207& &238&$239.4(7.3)$& & 
$\left\{\begin{array}{c}220(2)^*\cr 209(4)^{**}\end{array}\right.$\\
$f_{K^*}$ & 236& 267& 241& &241&$255.5(6.5)$& & $217(5)^\dag$\\
$f_\phi$ &245&336&259&&&$270.8(6.5)$& & $229(3)^\ddag$  
\end{tabular}
\end{ruledtabular}
\begin{flushleft}
${}^*$ derived from the experimental value for $\Gamma_{\rho^0\to
  e^+e^-}$.\\
${}^{**}$ derived from the experimental value for $\Gamma_{\tau\to\rho\nu_\tau}$.\\
${}^\dag$ derived from the experimental value for 
$\Gamma_{\tau\to K^*\nu_\tau}$.\\
${}^\ddag$ derived from the experimental value for $\Gamma_{\phi\to
  e^+e^-}$. 
\end{flushleft}

\end{table}

In Table~\ref{tab:dc} we present our predictions for the
light and heavy-light meson decay constants calculated using the meson
wave functions which were obtained as the numerical solutions of the
quasipotential equation.\footnote{We roughly estimate the uncertainties in our
calculations to be about ten MeV for light mesons and of a several
MeV for heavy-light mesons.} The values~\footnote{For the
evaluation of $f_M^{\rm NR}$ the relativistic wave functions were
used. Thus the difference of the
pseudoscalar and vector decay constants in this limit results from the
difference of the corresponding relativistic wave functions.} of
$f_M^{\rm NR}$, obtained from the nonrelativistic expression
(\ref{eq:fnr}), as well as the values of different contributions  
$f_M^{(1,2,3,4)}$ (\ref{eq:fpv1})--(\ref{eq:fpv4})
and the complete relativistic values of $f_M$ (\ref{eq:fpe})  are given. In
Table~\ref{tab:dce} we compare our results~\footnote{In our model 
$\rho$ and $\omega$ mesons are degenerate, therefore their decay
constants are equal. The experimental value
for the decay constant of the $\omega$ meson, derived from 
$\Gamma_{\omega\to e^+e^-}$ \cite{pdg}, is $f_\omega=195(3)$~MeV.}
for the decay constants $f_M$ of light mesons
with other quark model predictions  \cite{gi,mr,k,hjd}, recent values from
two- \cite{ak} and three-flavour \cite{milc} lattice QCD   and
available experimental data \cite{pdg}. It is clearly seen that the
nonrelativistic predictions are significantly overestimating all
decay constants, especially for the pion (almost by a factor of 10). The
partial account  of relativistic corrections by keeping in
Eq.~(\ref{eq:fpe}) only the first term $f_M^{(1)}$
(\ref{eq:fpv1}), which is usually used for semirelativistic
calculations,  does not substantially improve the situation. The
disagreement is still large.  This is connected with the anomalously
small masses of light pseudoscalar mesons exhibiting their chiral
nature. In the semirelativistic quark model \cite{gi} the pseudoscalar
meson mass is replaced by the so-called mock mass $\tilde M_P$, which is equal to
the mean total energy of free quarks in a meson, and with our wave
functions: $\tilde
M_\pi=2\langle\epsilon_q(p)\rangle\approx 1070$~MeV ($\sim 8 M_\pi$)
and $\tilde 
M_K=\langle\epsilon_q(p)\rangle+\langle\epsilon_s(p)\rangle\approx
1232$~MeV ($\sim 2.5 M_K$). Such replacement yields values of $f_P^{(1)}$ 
which are still $\approx 1.4$ times larger than experimental ones
(cf. \cite{gi}). As we see from Table~\ref{tab:dc}, 
it is not justified to neglect contributions of
the negative energy intermediate states for light meson decay
constants. Indeed, the values of $f_M^{(2+3)}+f_M^{(4)}$ are large and
negative (reaching $-76\%$ of $f_\pi^{(1)}$ for the pion) 
thus compensating the overestimation of decay constants by
the positive-energy contribution $f_M^{(1)}$. This is the consequence
of the smallness of the 
light pseudoscalar meson masses compared to the energies of their
constituents. The negative-energy contributions (\ref{eq:fpv3}),
(\ref{eq:fpv4}) are proportional to the ratio of the meson binding
energy $M-\epsilon_1(p)-\epsilon_2(p)$ to its mass and quark energies.
For mesons with 
heavy quarks this factor is small and leads to the suppression of negative-energy
contributions. This results in the dominance of the positive-energy
term $f_M^{(1)}$. Indeed the
negative-energy terms for heavy-light $D$ and $B$ mesons give
$10-15$\% contributions (see Table~\ref{tab:dc}) which have the
typical magnitude of the heavy quark corrections. This explains the
closeness of the obtained values of constants to our previous results
\cite{efgdc}.  
On the other hand, for light mesons,
especially for the pion and kaon, the binding energies are not small on the
meson mass and quark energy scales and, thus, such factor gives no
suppression. The complete relativistic expression (\ref{eq:fpe}) for decay
constants $f_M$  brings theoretical predictions for
light mesons in good
agreement with available experimental data. 

The comparison of our values of the decay constants of light mesons
with other predictions in 
Table~\ref{tab:dce} indicate that they are competitive even with the results of
more sophisticated approaches (e.g. \cite{mr}) which are based on the
Dyson-Schwinger and Bethe-Salpeter equations. On the other hand, our
model is more selfconsistent than some other approaches \cite{hjd,gi}. 
We calculate the meson wave functions by solving the
quasipotential equation in contrast to the models based on the
relativistic Hamilton dynamics \cite{hjd} where various ad hoc
wave function parameterizations are employed.

\begin{table}
  \caption{Pseudoscalar and vector decay constants of heavy mesons (in
    MeV).}
  \label{tab:dchm}
\begin{ruledtabular}
\begin{tabular}{ccccccccc}
Constant&\multicolumn{2}{l}{\underline{\hspace{.4cm}Quark models\hspace{.4cm}}}&
\multicolumn{2}{l}{\underline{\hspace{1.1cm}Lattice QCD\hspace{1.1cm}}} &
\multicolumn{3}{l}{\underline{\hspace{.8cm}QCD sum
    rules\hspace{.8cm}}}& \underline{\hspace{.3cm}Experiment\hspace{.3cm}}\\

&this work& \cite{ckwn} &\cite{ak} &
\cite{ml,hpqcd} &\cite{nar}& \cite{ps}& \cite{jl} & \cite{pdg,cleo}\\ 
\hline
$f_D$ & 234& 230(25) & 225(14)(40)&201(3)(17)&203(20)& 195(20)&
&$222.6(16.7)(^{2.8}_{3.4})$  \\
$f_{D_s}$ & 268& 248(27)& 267(13)(48)& 249(3)(16)&235(24)& & 
&266(32) \\
$f_{D_s}/f_D$ & 1.15&1.08(1)& &1.24(1)(7)  &1.15(4)& & & \\
$f_B$ & 189& 196(29)& 208(10)(29)& 216(9)(19)(6) &203(23)& 206(20)&210(19)&  \\
$f_{B_s}$ &218 &216(32)&250(10)(35)& 259(32)& 236(30)& &244(21)& \\
$f_{B_s}/f_B$ & 1.15 & 1.10(1)& & 1.20(3)(1)&1.16(4)& &1.16&    
\end{tabular}
\end{ruledtabular}

\end{table}

In Table~\ref{tab:dchm} we confront our results for pseudoscalar
decay constants of the heavy-light mesons as well as their ratios  with the
recent predictions based on  the Salpeter equation \cite{ckwn},
values from the unquenched two- \cite{ak} and three-flavour
\cite{ml,hpqcd} lattice QCD,\footnote{The recent quenched lattice QCD
values \cite{chllc} 
for the pseudoscalar decay constants are $f_K=152(6)(10)$ MeV, $f_D=235(8)(14)$ MeV
and $f_{D_s}=266(10)(18)$ MeV. }  QCD sum rules \cite{nar,ps,jl} and  available
experimental data \cite{pdg}.  Reliable experimental data, up till recently,
existed only for $f_{D_s}$, which was measured by several experimental
collaborations (ALEPH, DELPHI, L3, OPAL, Beatrice, CLEO, E653, WA75,
BES) both in the $D_s\to\mu\nu$ and the $D_s\to\tau\nu$ decay
channels. At present, experimental errors are still rather large for this
constant. Very recently, the CLEO Collaboration \cite{cleo} published a
relatively precise value for the decay constant $f_D$ measured in
$D\to\mu\nu$ decay.  We see from Table~\ref{tab:dchm} that there is a
good (within error bars) agreement between all presented theoretical
predictions as well as with available experimental data.

In summary, the weak decay constants of pseudoscalar and vector light
and heavy-light mesons were investigated with the special emphasize on the
role of relativistic effects. For our calculations we used the
meson wave functions which were obtained by the numerical solution of
the quasipotential equation with the nonperturbative treatment of all
spin-dependent and spin-independent relativistic contributions to the
quark interaction potential. It was argued that both positive-
and negative-energy 
parts of the quark propagators in the weak annihilation loop should be
taken into account. The positive-energy contributions, which are
usually considered in the semirelativistic quark models,
significantly overestimate the decay constants of light
mesons. We showed that the negative-energy contributions to the light
meson decay constants are large and negative. Their account is
necessary to bring theoretical predictions in  agreement
with experimental data. On the other hand, these negative-energy
contributions are considerably smaller for decay constants of
heavy-light mesons and have the order of magnitude of the lowest correction in
the heavy quark expansion. The consistent inclusion of relativistic
effects coming both from the quark propagators and the meson wave
functions considerably improve the accuracy and reliability of the
obtained predictions.

The authors are grateful to A. Ali Khan, M. M\"uller-Preussker and
V. Savrin  for support and useful discussions.  Two of us
(R.N.F. and V.O.G.)  were supported in part by the {\it Deutsche
Forschungsgemeinschaft} under contract Eb 139/2-3 and by the {\it Russian
Foundation for Basic Research} under Grant No.05-02-16243.

\end{document}